# Decomposition of La$_{2-x}$Sr$_x$CuO$_4$ into Several La$_2$O$_3$ Phases at Elevated Temperatures in Ultra-High Vacuum inside a Transmission Electron Microscope

Jong Seok Jeong[1,*], Wangzhou Wu[1], Mehmet Topsakal[2], Guichuan Yu[3], Takao Sasagawa[4], Martin Greven[3], K. Andre Mkhoyan[1,*]

[1] Department of Chemical Engineering and Materials Science, University of Minnesota, Minneapolis, Minnesota 55455, United States

[2] Center for Functional Nanomaterials, Brookhaven National Lab, 735 Brookhaven Ave, Upton, New York 11973, United States

[3] School of Physics and Astronomy, University of Minnesota, Minneapolis, Minnesota 55455, United States

[4] Materials and Structures Laboratory, Tokyo Institute of Technology, Kanagawa 226-8503, Japan

[*] Corresponding Authors: jsjeong@umn.edu, mkhoyan@umn.edu

**ABSTRACT:**

We report the decomposition of La$_{2-x}$Sr$_x$CuO$_4$ into La$_2$O$_3$ and Cu nanoparticles in ultra-high vacuum, observed by *in-situ* heating experiments in a transmission electron microscope (TEM). The analysis of electron diffraction data reveals that the phase decomposition process starts at about 150 °C and is considerably expedited in the temperature range of 350-450 °C. Two major resultant solid phases are identified as metallic Cu and La$_2$O$_3$ by electron diffraction, simulation, and electron energy-loss spectroscopy (EELS) analyses. With the aid of calculations, La$_2$O$_3$ phases are further identified to be derivatives of a fluorite structure—fluorite, pyrochlore, and (distorted) bixbyite—characterized by different oxygen-vacancy order. Additionally, the bulk plasmon energy and the fine structures of the O $K$ and La $M_{4,5}$ EELS edges are reported for these structures, along with simulated O $K$ X-ray absorption near-edge structure. The resultant Cu nanoparticles and La$_2$O$_3$ phases remain unchanged after cooling to room temperature.

Ever since the discovery of superconductivity below about 4 K in mercury a century ago, research has aimed to achieve superconductivity at high temperatures. After the 1986 discovery of high-temperature superconductivity in copper-oxide ceramics such as $La_{2-x}Sr_xCuO_4$, these lamellar oxides quickly became one of the most heavily studied materials families because of their record high superconducting transition temperature at ambient pressure ($T_c$=135 K), mysterious superconducting mechanism, and numerous other intriguing strong-correlation phenomena [1]. Charge-carrier doping of the quintessential $CuO_2$ planes is typically achieved either through changes in the density of oxygen interstitials or via cation substitution (as in $La_{2-x}Sr_xCuO_4$) [2]. More recently, carrier doping has also been demonstrated electrostatically [3]. One obstacle toward applications of these and other oxides, including systems in which the goal is to electrostatically modify surface charge-carrier density, is their chemical instability toward the formation of oxygen vacancies [3-7]. Even though the $T_c$ and hence the operating temperature of these materials is still well below room temperature (RT), it is highly desirable to understand their instability in low oxygen partial pressure at elevated temperatures, which might aide the preparation of high-quality materials that feature less disorder.

Here we report the detailed analysis of the results of *in-situ* heating experiments for bulk $La_{2-x}Sr_xCuO_4$, one of the most widely-studied cuprate superconductors [8], at the non-superconducting doping level $x$=0.03. We analyze the phase separation/decomposition phenomenon under the ultra-high vacuum (UHV) conditions of a transmission electron microscope (TEM). This analysis involves both *in-situ* TEM and electron energy-loss spectroscopy (EELS) data obtained during sample heating from RT to 700 °C (973 K), and then after cooling back to RT.

It should be noted that Petrov et al. [9] previously reported an investigation of the thermodynamic stability of several ternary oxides, including La-Cu-O, at elevated temperatures (973-1573 K range) and under oxygen partial pressures of $10^{-15}$–1 atm by analyzing electromitove force (emf) vs temperature data, and that Gao et al. [10] reported the observation of partial decomposition of $La_{1.867}Th_{0.100}CuO_{4.005}$ into $LaCuO_2$, $La_2O_3$, CuO, and a very small amount of Cu at 500 °C using X-ray diffraction and thermal gravimetric analysis. However, detailed studies of critical local nano-scale processes during such chemical decomposition and of the resultant phases in these oxides are still lacking. Therefore, the detailed structural investigation of oxide superconductors at elevated temperatures remains of considerable significance [11-13]. TEM equipped with EELS enables the observation of the structural and chemical evolution of a material during heating/cooling in a rather unique manner with high spatial resolution, which is essential for such a study.

The single-crystal $La_{2-x}Sr_xCuO_4$ ($x$=0.03) studied here was grown by the traveling-solvent floating-zone technique [14,15]. Cross-sectional TEM specimens were prepared using a focused ion beam (FEI Quanta 200 3D and FEI Helios NanoLab G4) using 30 kV Ga ions, followed by 1–5 kV milling to minimize damaged surface layers. To avoid the carbon contamination build-up during imaging and EELS measurements, the TEM specimens were treated with standard plasma exposure using Fischione Plasma Cleaner (Model 1020). High-resolution scanning TEM (STEM) images were obtained with an aberration-corrected FEI Titan G2 60-300 STEM, operated at 300 keV. High-angle annular dark-field (HAADF) and low-angle annular dark-field (LAADF) STEM images were recorded with detector angles of 41–200 mrad and 10–41 mrad, respectively. The convergence semi-angle of the STEM incident probe was 24.3 mrad.

The heated-stage TEM experiment was performed using a Gatan 652 double-tilt heating holder in an FEI Tecnai G2 F30 STEM with TWIN pole-piece operated at 300 keV and equipped with a Gatan 4k×4k Ultrascan CCD. Bright-field TEM (BF-TEM) images and selected-area electron diffraction (SAED) patterns were acquired at each temperature setting. EELS spectra were recorded using a Gatan Image Filter (GIF) spectrometer attached to this microscope. The EELS data were acquired by using a selected-area aperture (about 250 nm in diameter) on a fixed position of the specimen during the heating experiment. During these measurements, the temperature of the specimen was ramped up at a rate of 30–50 °C/min, starting from 20 °C. At each temperature of interest, the TEM and EELS data were acquired after about 5 min of thermalization.

Core-electron excitation calculations for comparison with measured EELS data were performed from the first-principle calculations by solving the Bethe-Salpeter equation (BSE) using the EXCITING code, within the framework of the full-potential linearized augmented plane-wave (FLAPW) method [16,17]. Experimental crystal information was used for known structures—$La_2CuO_4$ and cubic bixbyite $La_2O_3$ ($c$-$La_2O_3$). Supercells of defective fluorite ($f$-$La_2O_3$) and defective pyrochlore ($p$-$La_2O_3$) structures were created by randomly placing oxygen vacancies in standard 2×2×2 fluorite ($AX_2$) and 1×1×1 pyrochlore ($A_2B_2X_7$) structures. For the $p$-$La_2O_3$ structure, La was used for both A and B cations. Distorted bixbyite $La_2O_3$ ($b$-$La_2O_3$) structure was created by adapting structural information from similar distorted bixbyite α-$Mn_2O_3$ [18,19]. The supercells of $f$-$La_2O_3$, $p$-$La_2O_3$, $b$-$La_2O_3$, and $c$-$La_2O_3$ include 32 La and 48 O atoms. The supercell of $c$-$La_2O_3$, which has higher symmetry, was further reduced to a primitive cell with 16 La and 24 O atoms for the calculation. The ground-state computations were performed with GGA-PBEsol functional [20] and the detailed description of both ground-state and BSE calculations can be found in the Supplemental Material [21].

Figure 1 shows a representative low-magnification BF-TEM image and a high-resolution HAADF-STEM image of the $La_{2-x}Sr_xCuO_4$ crystal at RT, before heating, viewed along [110]. $La_{2-x}Sr_xCuO_4$ features a structural transition from orthorhombic at low temperature (LTO phase) to tetragonal at high temperature (HTT phase), with a transition temperature ($T_{st}$) that decrease approximately linearly with doping [8]: $T_{st}$~530 K for $x$=0 and $T_{st}$~465 K for $x$=0.03. Therefore, the initial $La_{2-x}Sr_xCuO_4$ specimens studied in this work, is in the LTO phase at RT. The SAED pattern, shown in the inset of Fig. 1(a) and indexed using a distorted $K_2NiF_4$ (*C*mca) structure [22,23], confirms the single-crystalline nature of the specimen in the LTO phase. The atomic-resolution HAADF-STEM image in Fig. 1(b) also shows the $K_2NiF_4$ structure. In the BF-TEM image, mainly two different contrasts were observed: a sharp contrast along <113> and a broad contrast along <110>. The former originates from antiphase boundaries in the ($\bar{1}$13) and (1$\bar{1}$3) planes, which are parallel to the electron beam. An example of these antiphase boundaries is shown in the upper inset of Fig. 1(b). The latter also originates from the antiphase boundaries in the (113) plane, which is crystallographically almost equivalent to the ($\bar{1}$13) and (1$\bar{1}$3) planes, but inclined to the electron beam direction, and thus it appears broad.

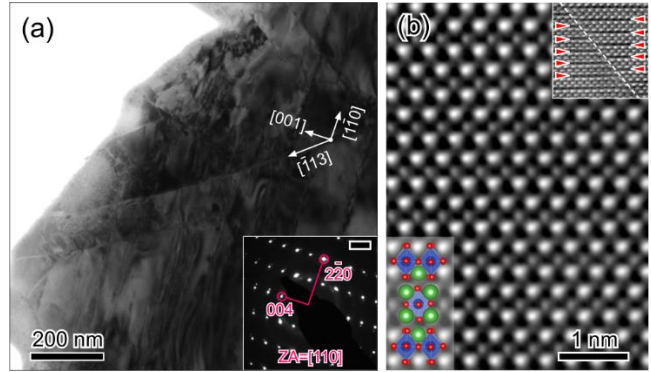

**Figure 1.** Low-magnification BF-TEM image (**a**) and high-resolution HAADF-STEM image (**b**) from a [110]-oriented $La_{2-x}Sr_xCuO_4$ ($x$=0.03) specimen. The inset to (**a**) is a corresponding SAED pattern. The scale bar of the inset is 2 nm$^{-1}$. The atomic-resolution HAADF-STEM image shows the $K_2NiF_4$ structure, which is the structure of the $n$=1 Ruddlesden-Popper phase with stacking sequence $(La^{3+}Cu^{2+}O_3)^{1-}$–$(La^{3+}O^{2-})^{1+}$ along the *c*-axis. A schematic of the atomic structure is superimposed in lower-left corner of panel (**b**): La/Sr (green), Cu (blue), and O (red). LAADF-

STEM image of an {113}-antiphase boundary is shown in the upper-right inset of the panel (**b**).

During the heating of the La$_{2-x}$Sr$_x$CuO$_4$ crystal inside the microscope, we found that the specimen undergoes phase separation/decomposition, which is different from the LTO-to-HTT phase change. Figure 2 shows series of BF-TEM images and corresponding SAED patterns at different temperatures going from 20 °C to 700°C. Extra diffraction spots start to appear in the SAED patterns at 150 °C (indicated by the red arrows in Fig. 2). Some of these diffraction spots become more obvious at 500 °C (indicated by the yellow rhombus), as nanoparticles appear in TEM images. The nanoparticles were identified as metallic Cu from core-loss EELS data (inset). The diffraction pattern from the La$_{2-x}$Sr$_x$CuO$_4$ crystal (the red parallelogram) disappears completely at 700 °C and only the new diffraction pattern (the yellow rhombus) remains while the Cu nanoparticles grow bigger. This result implies that, during heating, Cu is depleted from the La$_{2-x}$Sr$_x$CuO$_4$ matrix to form the nanoparticles, and thus a new phase without Cu forms. It appears that Cu diffuses out of the matrix to nucleate particles starting at about 150 °C, accompanied by the formation of a new phase in a small portion of the matrix. However, the Cu nanoparticles start to be clearly visible in BF-TEM images only after 380 °C (see Video 1, Supplemental Material [21]), mainly because smaller-sized Cu particles are invisible with the complex contrast of BF-TEM images. The initial diffusion of Cu atoms occurs at about 150 °C (or 423 K), which is very close to the temperature of the LTO-to-HTT phase change ($T_{st}$~465 K). However, the agreement of these two temperatures would seem coincidental, since the chemical decomposition temperature can be expected to be a function of the oxygen partial pressure [24], which is extremely low in our UHV experiment. We note that quantitative, reproducible charge transport data have been obtained for La$_{2-x}$Sr$_x$CuO$_4$ up to about 1000 K in a suitable atmosphere, which constitutes indirect evidence for the absence of chemical decomposition at elevated oxygen partial pressure [25]. At temperatures not far above RT, the diffusion coefficient is small and, as a result, the growth of nanoparticles is limited. At elevated temperatures of about 350 °C, the diffusion coefficient of the Cu atoms dramatically increases, as $D = D_0 exp\{-Q_d/k_B T\}$, where $D_0$ is the diffusion constant and $Q_d$ is the activation energy, resulting in accelerated growth of Cu particles observed in our experiment. Additionally, because the oxygen content slowly changes upon heating, and quickly decreases above 250 °C (Fig. S3 [21]), the phase decomposition will also be affected due to changes in stoichiometry at the elevated temperature.

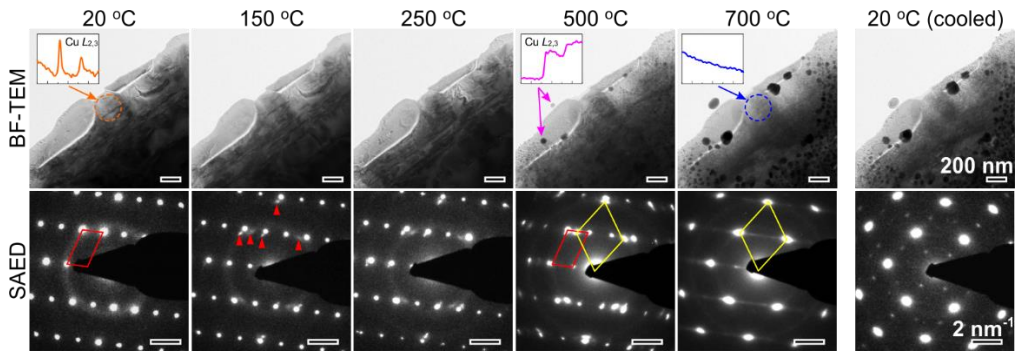

**Figure 2.** BF-TEM images (top panels) and corresponding SAED patterns (bottom panels) at different temperatures during a heating experiment, and after cooling to 20 °C. Insets to panels at 20 °C, 500 °C, and 700 °C are core-loss EELS data in the range of 910–970 eV (the arrows indicate the areas of EELS acquisitions). The SAED pattern at 20 °C is the same [110] pattern (indicated by the red parallelogram) as that shown in Fig. 1(a). The red arrows at 150 °C show extra diffraction spots from the new phase formation. The yellow rhombus at 700 °C shows the completion of the phase separation.

The reason why metal Cu nanoparticles rather than a stable $CuO_2$ form in the experiment is likely the result of the extremely low oxygen partial pressure. The vacuum level in the pole-piece area of the TEM, where the sample was located during the experiment, is about $10^{-8}$–$10^{-7}$ Torr [26]. For this vacuum level and at such elevated temperatures, copper-oxide is expected to dissociate into metallic Cu and $O_2$ gas [27], which is consistent with our observations.

Once the chemical phase separation was completed, Cu was not detected in the new phase, as shown in the EELS data in Fig. 2, which implies that some form of lanthanum/strontium-oxide should be expected. The crystal structure of this new phase, after the Cu dissociation, was determined using SAED pattern analysis, complemented by diffraction pattern simulations and EELS analysis. Many SAED patterns from different zone-axes of the crystal were recorded to avoid ambiguity of the structure determination (Fig. S4 [21]). Because EELS compositional analysis revealed that the O/La atomic ratio changes approximately from 2 in $La_{2-x}Sr_xCuO_4$ to 1.5 in the new phase (Fig. S3 [21]), compositionally the new phase should be $La_2O_3$. Note that it is not clear whether Sr incorporates into the new phases or forms an additional phase such as SrO, because in our compositional analysis, Sr with such a low doping level ($x$=0.03) was not detectable. $La_2O_3$ is the most common form of lanthanum-oxide and has a hexagonal RT structure [28,29]. However, hexagonal $La_2O_3$ diffraction patterns were not found to match the measured patterns. On the other hand, a cubic bixbyite structure (also known as C-rare earth structure), which is usually found in sesquioxide (e.g. $Y_2O_3$, $Dy_2O_3$, $In_2O_3$, $Sm_2O_3$, etc.) and is stable at higher temperatures [28,29], can produce such diffraction patterns.

Three experimental SAED patterns from different areas of the sample measured at 700 °C are shown in Fig. 3. Because the cubic bixbyite structure (space group: $I$a-3) belongs to the bcc lattice, the 111 diffraction is forbidden, whereas such diffraction spots were visible in our experiment (see Fig. 3(b)). Geller and Norrestam [18,19] reported that it is possible to have a distorted bixbyite structure with orthorhombic symmetry (space group: $P$bca) in α-$Mn_2O_3$. If the $La_2O_3$ phase in our experiments is slightly distorted with $LaO_6$ octahedra tilts, and has a symmetry lower than cubic, the forbidden 111 diffraction spots would be allowed. A phase decomposition, where $Nd_{2-x}Ce_xCuO_{4\pm\delta}$ decomposes into strained bixbyite $(Nd,Ce)_2O_3$ phase, has been observed before [30]. In order to test this possibility, we adapted the structure parameters of α-$Mn_2O_3$ and simulated dynamical electron diffraction patterns using *Multislice* code (Fig. S4 [21]) [31]. The results suggest that, indeed, the distorted bixbyite structure will form diffraction patterns with more visible spots. However, as shown Fig. S4 [21], most diffraction patterns observed in this study are much simpler than the ones from the bixbyite structure, whether distorted or not, which indicates that most of the resultant structures are close to the bixbyite structure, but with simpler atomic structures (*e.g.*, fluorite or pyrochlore). Such structures were observed in oxides such as (1-$x$)ZrO$_2$·$x$$Ln$O$_{1.5}$ ($Ln$ = La to Gd) [32]. To the best of our knowledge, however, the observation of fluorite and pyrochlore structures in lanthanum oxides has not yet been reported.

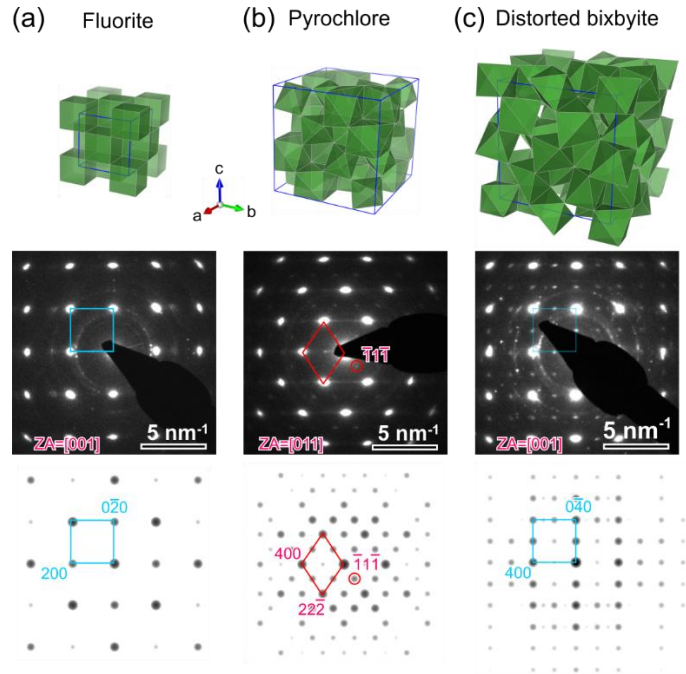

**Figure 3.** Polyhedral models of fluorite (**a**), pyrochlore (**b**), and distorted bixbyite (**c**) structures (top). The coordination of La atoms in fluorite and bixbyite structures is 8 and 6, respectively, and the pyrochlore structure has a mixture of both. Examples of experimental SAED patterns from different areas of the sample that can be uniquely assigned for fluorite, pyrochlore, and distorted bixbyite structures are presented (middle). Simulated diffraction patterns from these structures are shown for comparison (bottom). The diffraction patterns were simulated for a sample thickness of 30 nm. In (**b**), the 111 diffraction spot, which is forbidden in the cubic bixbyite and fluorite structures, is visible. In (**c**), satellite spots, which are forbidden for fluorite, pyrochlore, and cubic bixbyite structures, are visible, indicating that a distorted bixbyite structure is present (also see Fig. S5 [21]).

The bixbyite structure is a derivative of CaF$_2$-type fluorite structure (space group: $F$m-3m) with anion deficiency. Combining 2×2×2 oxygen deficient unit cells of the fluorite structure in an appropriate manner results in a unit cell of the bixbyite structure [33]. The pyrochlore structure (space group: $F$d-3m) is another fluorite-based structure with an oxygen deficiency. In the pyrochlore structure, two differently oxygen deficient unit cells (with two different cations) of the fluorite structure are brought together to form a three-dimensional checker-board-like $2\times2\times2$ supercell [33]. The anion/cation atomic ratios in fluorite, pyrochlore, and bixbyite structure are 2, 1.75, and 1.5, respectively. Because the O/La atomic ratio varies from 2 to 1.5 during our heating experiments, all these structures appear to be feasible compositionally. However, the O/La ratio change is mainly due to the change in relative volume fraction between $La_{2-x}Sr_xCuO_4$ and $La_2O_3$ (regardless of which structure $La_2O_3$ has).

To unambiguously identify the structures of the resultant phases after decomposition, SAED patterns from several different zone axes were acquired at 700 °C and after cooling to see if the phases remain at RT. Experimentally measured SAED patterns and simulated diffraction patterns from four different structures—fluorite, pyrochlore, cubic bixbyite, and distorted bixbyite—are compared in Fig. S4 [21]. Because pyrochlore and bixbyite structures are derivatives from the fluorite structure, they share the same overall diffraction pattern; the lower the structural symmetry, the more satellite spots should be visible. The data show not only the main spots of the fluorite structure, but also satellite spots. The diffraction analysis reveals that all four structures appear to co-exist at 700 °C and after cooling. Although it is not clear how these structures are distributed throughout the sample, in some regions of the sample, only one structure is dominant whereas in others they are mixed. The SAED patterns in Fig. 3 are examples of these cases. The SAED pattern along [001] in Fig. 3(c) proves that the distorted bixbyite also exists, as expected (see also Fig. S5 [21]).

Because all fluorite-based structures exist at 700 °C and the O/La ratio at that temperature is about 1.5, one can imagine that the fluorite and pyrochlore structures should have extra oxygen vacancies, *i.e.*, that they are defective structures. In non-defective cases, from fluorite to pyrochlore to bixbyite, the oxygen deficiency increases with distinct distributions of oxygen vacancies in the unit cells. At 700 °C in our work, the O/La ratio in all structures is approximately 1.5, and thus only oxygen vacancy distribution can differentiate between defective fluorite, defective pyrochlore, and bixbyite structures. If the positions of the oxygen vacancies are completely random, *i.e.*, all oxygen sites are partially occupied, the structure becomes the defective fluorite. In other words, the structural transitions between defective fluorite to defective pyrochlore to bixbyite could be considered as changes in the ordering of oxygen vacancies. The slight discrepancy between the experimental and simulated diffraction patterns might be due to small local variations in the degree of the oxygen vacancy order.

We observed nanoscale grain contrast in high-magnification BF-TEM images (Fig. S6 [21]), which reveals the presence of nanometer-size grains in the sample. Regions selected by the selected-area aperture (~250 nm in diameter) were found to include several grains and more than one phase. Despite the existence of grains, the region in the aperture mostly features the same crystallographic orientation, probably because all resultant phases have fluorite-based structures. The grains in the region might have the same crystallographic orientation, but different degrees of oxygen vacancy order. All these observations support the diffraction-based analysis discussed above. Local variations of oxygen vacancies during heating are also evident from the strain contrast changes in BF-TEM images (see Videos 2 and 3, Supplemental Material [21]).

Electronic structure evolution during the phase decomposition was monitored using EELS. A low-loss region for the plasmon peak and core-loss regions for the O $K$, La $M_{4,5}$, and La $O_{2,3}$ edges were investigated as the temperature was increased. The results are shown in Fig. 4. Even though subtle changes in the SAED patterns were observed upon heating to 150 °C, obvious changes in the EELS data were observed between 300-450 °C, as the majority of the phase decomposition and Cu nanoparticle formation was observed in this temperature range (see Video 1, Supplemental Material [21]). As can be seen in Fig. 4(a), the bulk plasmon peak was found to shift from 25.2 eV in $La_{2-x}Sr_xCuO_4$ to 24.6 eV in $La_2O_3$, whereas no detectable change was observed in the La $O_{2,3}$ edge fine structure. From a simple free-electron model [34], the bulk plasmon energy is estimated to be 24 eV in $La_{2-x}Sr_xCuO_4$ and 19 eV in $La_2O_3$. While this model is too simple to be quantitatively accurate, it correctly predicts the shift of the bulk plasmon peak to lower energy-loss.

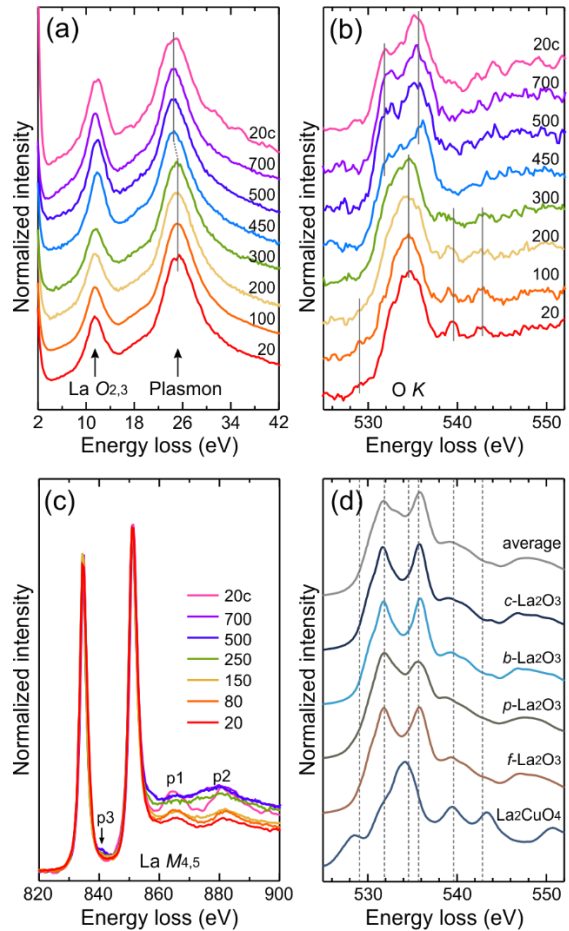

**Figure 4.** (**a**) Low-loss, (**b**) O $K$ edge, and (**c**) La $M_{4,5}$ edge EELS spectra at selected temperatures (in °C) during a heating experiment. The label '20c' represent the 20 °C after cooling. (**d**) Calculated O $K$ edge XANES spectra for $La_2CuO_4$, $f$-$La_2O_3$, $p$-$La_2O_3$, distorted bixbyite $La_2O_3$ ($b$-$La_2O_3$), and cubic bixbyite $La_2O_3$ ($c$-$La_2O_3$). An average spectrum of the four $La_2O_3$ structures is also presented. The calculated spectra were aligned to the experimental spectra and the major peak positions in (**b**) are demonstrated in (**c**) by the dotted lines for comparison.

The changes in the O $K$ edge fine structure are more significant. As shown in Fig. 4(b), in the 300-450 °C temperature range, the main peak of the O $K$ edge at 534.6 eV splits into two peaks at 532.0 and 535.8 eV, and the two minor peaks at 539.1 and 542.8 eV are suppressed. These changes in the fine structure of the O $K$ edge agree with the O $K$ edges calculated, based on the structures of the new phases discussed above. Because in a real crystal the level of crystal relaxation may vary spatially, for each system with one particular oxygen vacancy distribution case, we performed the calculations with (Fig. 4(d)) and without (Fig. S7 [21]) crystal relaxation. The relaxed structures of defective fluorite/pyrochlore and bixbyite are very similar because they have the same number of atoms in the unit cell, and only differ with regard to the location of oxygen vacancies (Fig. S2 [21]). Therefore, not surprisingly, the calculated X-ray absorption near-edge structure (XANES) spectra from all four structures are very similar. It should be noted that in these kinds of calculation it is impractical to properly incorporate a partial oxygen-site occupancy, which is necessary to hold the $F$m-3m or $F$d-3m symmetry of the systems, as seen in experimental SAED patterns. This limitation prevents the direct comparison of our calculations with the experimental results. Nevertheless, the calculated O $K$ edge spectra are in good agreement with the data (especially the spectrum obtained upon averaging over those for all four structures—see Fig. 4(d)), which provides additional support for the diffraction-based analysis of the decomposed $La_2O_3$ phases.

The La $M_{4,5}$ edge also shows subtle, yet detectable changes upon going from $La_{2-x}Sr_xCuO_4$ to $La_2O_3$. As shown in Fig. 4(c), the peak labeled by 'p2' shifts to the lower energy-loss value, and the intensity ratio of peaks 'p1' and 'p2' changes upon heating (see Fig. S8 [21]). Additionally, a new small shoulder 'p3' appears after the phase change. No significant change in the sharp white lines was observed. We note that the calculation for reliable La $M_{4,5}$ edge in the supercell size, used in this work, is impractical and its interpretation is not straightforward because it needs much complicated consideration of multi-electron excitation and spin-orbit effects in calculations, which are beyond of the scope of this work. We only report data showing changes in the La $M_{4,5}$ fine structure after phase decomposition for later studies.

To test if the observed phase changes are reversible under our UHV experimental conditions or if the new phases formed by heating remain unchanged after cooling, the sample was cooled to RT inside the TEM, and all measurements were repeated. No significant changes were detected in the SAED pattern (Fig. 2 and Fig. S4 [21]) as well as in EELS data (Fig. 4), which indicates that the process is not reversible and that the defective fluorite/pyrochlore phases remain upon cooling. The only noticeable difference in the SAED patterns of Fig. 2 before and after cooling is that the pattern at 700 °C features streaks, which are not often observed (Fig. S4 [21]). These streaks can be attributed to planar defects or interfaces, perpendicular to the streaks, in the area where the SAED pattern is acquired. The streaks disappear after cooling, and thus we do not completely rule out the possibility of further phase changes during cooling. More importantly, the resultant phases were observed throughout the sample after cooling, which suggest that it might be possible to synthesize such unique phases in the future. Further investigation of the long-term stability of these phases is needed.

In conclusion, we have reported on the chemical phase decomposition of $La_{2-x}Sr_xCuO_4$ into $La_2O_3$ and Cu nanoparticles, observed by *in-situ* heating experiments in TEM. The phase separation starts at about 150 °C and is expedited in the elevated temperature range of 350-450 °C. By electron diffraction, simulations, and EELS analyses, the resultant solid phases were identified as metallic Cu and fluorite-based structures—defective fluorite/pyrochlore and (distorted) bixbyite—characterized by different oxygen-vacancy order. EELS measurements demonstrated that a bulk plasmon peak, O $K$, and La $M_{4,5}$ edges can be used to monitor the phase changes. This phase decomposition is irreversible under the UHV conditions of TEM, and the resultant Cu nanoparticles as well as $La_2O_3$ remain unchanged after cooling to RT. We hope that these observations will allow a better understanding of the stability of ternary oxides and aide the advanced processing of superconducting oxides.

**Supplemental Material:**

See Supplemental Material for the detailed description of core-excitation calculations, the experimental and simulated electron diffraction data, additional TEM and EELS data, and the videos showing structural changes during heating.

**Acknowledgements:**

This work was supported in part by the NSF MRSEC under award number DMR-1420013, also in part by the Grant-in-Aid program of the University of Minnesota. STEM analysis was performed in the Characterization Facility of the University of Minnesota, which receives partial support from the NSF through the MRSEC. The XANES simulations were performed in the Minnesota Supercomputing Institute (MSI) at the University of Minnesota. TS was supported by a JST-CREST project [JPMJCR16F2] and a JSPS Grants-in-Aid for Scientific Research (B) [16H03847]. MT was supported by the LDRD grant at the Brookhaven National Laboratory (Grant No. 16-039) and Center for Functional Nanomaterials, which is a U.S. DOE Office of Science Facility, at Brookhaven National Laboratory under Contract No. DE-SC0012704.